\documentclass[12pt]{article}
\usepackage{graphicx,xr}
\usepackage{amsmath}
\usepackage{bbm}
\usepackage{bm}

\newcommand{\nvec}[1]{{\mathbf #1}}
\newcommand{\tmop}[1]{\ensuremath{\operatorname{#1}}}

\title{Exploring molecular dynamics with forces from n-body potentials using MATLAB}
\author{Suhail Lubbad and Ladislav Kocbach \\
 {\small Dept. of Physics and Technology, University of Bergen, Allegaten 55,
          5007 Bergen, Norway}} 
\date{}

\begin{document}
\maketitle
\begin{abstract}
   Molecular dynamics is usually performed using specialized software packages. We present
   methods to perform exploratory studies of various aspects of molecular dynamics
   using MATLAB. Such methods are not suitable for large scale applications, but they can
   be used for developement and testing of new types of interactions and other 
   aspects of the simulations,  or simply for illustration of the 
   principles for instruction and education purposes. One advantage of MATLAB is the integrated
   visualization environment.
   We also present in detail exploration of forces obtained from 3-body potentials in 
   Molecular Dynamics in this framework. The methods are based on use of matrices and 
   multidimensional arrays for which MATLAB has a set of both linear algebra based
   as well as element-wise operations.  
   The numerical computations are based on these operations on the array objects rather than 
   on the traditional loop-based approach to arrays . The discussed  approach has a 
   straightforward implementation for pair potentials. 
   Applications to three-body interactions are the main aspect of this work, 
   but extension to any general form of n-body interactions is also discussed. 
   The methods discussed can be also applied without any change using the latest 
   versions of the package GNU OCTAVE as a replacement for MATLAB. The code examples 
   are listed in some detail, a full package of the MATLAB and OCTAVE codes is available
   for download. 
\end{abstract}
\section{Introduction}
   
Modelling of atomic motions in terms of classical Newtonian equations is 
usually called molecular dynamics (MD). Such modelling is considered useful
in many areas of material research, chemistry, molecular biology, with various
justifications for the applied approximations. In most applications large ensembles 
of atoms are treated using specialized software. In this paper we describe 
an alternative approach and describe 
methods which can be useful in design and exploration of new interactions, or
simply in education. While the large systems require optimization of the computations
and often possibility of parallel computation, the presented MD models
are supposed to be of the size up to tenths or possibly hundreds
of particles. We show that for such relatively small systems of model atoms
the computations can be carried out inside the systems MATLAB or OCTAVE. These systems 
should be relatively well known, but we give detailed references and small
presentation of features relevant for this work in the  Appendix \ref{AppendixA},

It is generally well known that MATLAB and related systems have interpreted
programming languages and this can make execution of large programs 
rather slow.
On the other hand, they have the ability to perform mathematical 
operations on complicated structured objects 
in one single statement using ordinary mathematical notation. This
property is essential for our approach. A simple
MD simulation code with pair potentials can be written in about
30 statements. These can be complemented by 
another set of routines of a comparable size (about 20 to 30 statements)
for quite powerful real time visualizations.
See   Appendix \ref{AppendixA}   for some special aspects in MATLAB language.

   The system MATLAB and its open source relative GNU OCTAVE have emerged 
as "matrix laboratory", i.e. a system for working with matrices. In their
latest versions they have been extended to handle not only matrices and 
vectors, but also multidimensional arrays. The systems are based on 
the use of an internal programming language. 
The coding is made easier by an internal on-line manual 
system, which explains in detail the various built in functions. The
disadvantage of computing with MATLAB is that the language is interpreted
and not compiled, thus not efficient in large scale calculations. On the 
other hand, the internal operations are based on optimized FORTRAN and C 
libraries and thus perform very well and very efficiently. Most of the 
statements about MATLAB are valid  to OCTAVE, except the graphical 
cabilities. MATLAB has a very advanced graphical system, its counterpart in 
OCTAVE has long been slightly different and much simpler, but in the latest versions
it is attempted to become very compatible.

   For exploratory work and small projects the fact that the programs 
are interpreted and not compiled is of a great advantage. Codes can be 
quickly adapted to any  new required tasks and tested in matter of seconds, 
since no compilation and linking cycles are required. 
The codes discussed in this paper and additional material are available for 
download \cite{DOWNLOAD}.
%
%
\section{Molecular dynamics and Verlet algorithm}
Generally, molecular dynamics refers to classical mechanics modeling of atomic
systems where the atoms are represented by mass points interacting with their 
surroundings, including the other atoms, by forces which can be represented by potentials,
schematically, for each atom there is a Newton equation
\begin{equation}
m_i \ddot{\nvec{r}_i}=-\nabla V(....,{\nvec{r}_i}, .... ) 
\label{Newton_1} 
\end{equation}
In many cases the coupled second order equations are solved using
the well known Verlet algorithm. This is a special difference scheme
for solving Newton-type equations
\begin{equation}
     \frac{d^2}{d t ^2}\ r(t)  \ =\ a (t) 
\label{diff_equation_1} 
\end{equation}
using the following difference scheme
\begin{equation}
r(t_{n+1})=2r(t_n)-r(t_{n-1})+a(t_n)\Delta t^2+O(\Delta t^4)
\label{Verlet_algorithm} 
\end{equation}
We illustrate the use by the code for one particle moving in attractive 
Coulomb potential, with the possibility of changing it to the so called soft
Coulomb, where the 1/r form is modified.
Such code is very popular for teaching demonstrations, but in our context it
is useful to test the accuracy and stability of the solutions in this 
very simple case. The code (we leave out the simple input and definition
statements) is
\parskip 0cm 
\parindent 0cm
{  \small  \begin{verbatim}
 1. t=0:tmax/ntime:tmax;
 2. r=zeros(2,ntime);
 3.     r(:,1)= r0;
 4.     r(:,2)= r1 ;
 5. for nt = 2:ntime 
 6.     Rt=r(:,nt);  accel=-Cm * Rt/sqrt(a^2+sum(Rt.^2))^3;
 7.    r(:,nt+1)= 2*r(:,nt)- r(:,nt-1) + accel *( t(nt)-t(nt-1) )^2;  
 8. end
\end{verbatim}        }
Here in line 6 we see that the vector Rt representing $\nvec{r}(t)$ is 
assigned the latest known
\parskip 0cm 
\parindent 0.95cm
position and the accelerations vector is evaluated according to
$$
\nvec{a}= - C_m \frac{\nvec{r}}{\left(\sqrt{a^2+r^2}\right)^3} 
= - C_m \nabla  \frac{1} { \sqrt{a^2+r^2} }
$$
Line 7 implements the Verlet algorithm as given above in eq.
\ref{Verlet_algorithm}. This line of code with only occasional small modifications
is used in all the discussed applications. Note that this operation
is performed on a vector. Later, when we consider more than one particle,
the symbol r can represent a "vector of vectors", then this line will 
become
\parskip 0cm 
\parindent 0cm
{  \small  \begin{verbatim}
  r(:,:,nt+1)= 2*r(:,:,nt)- r(:,:,nt-1) + accel *( t(nt)-t(nt-1) )^2;
\end{verbatim}        }
Similar notation is available in 
\parskip 0cm 
\parindent 0.95cm
the new versions of FORTRAN.

The simple code for this classical central force problem is 
very illustrative and useful for exploring the stability and
precission of the numerical method.
When the parameter a is zero, we have the Coulomb-Kepler problem and
the motion is periodic along an ellipse. If the a is different from
zero, the motion is generally not periodic and we get a rosette-like motion. 
This  elementary example
is included because the codes in this paper use the same Verlet implementation.


\section{Two body interactions}

The two-body forces considered here are forces between mass points, so that the
potential energies can only be functions of the distances between the particles, 
perhaps with addition of global forces of the environment on each particle, 
but we shall limit this discussion to the pair forces only, a global force 
is analogue to the first example of planetary motion above. 

The total potential energy is given by
\begin{equation}
V_{tot}(  \nvec{r}_1, ..... , \nvec{r}_N ) 
                = \sum_{i<j}
                    V\left( \left| \nvec{r}_1 - \nvec{r}_N  \right| \right)
\label{two_body_1} 
\end{equation}
and the force on each particle is obtained by gradient operation
\begin{equation}
m_k \ddot{\nvec{r}_k}  =-\nabla_k V_{tot}( \nvec{r}_1, ..... , \nvec{r}_N)
 =
 \sum_{j=1, j\ne k}  \left.  
                         \frac{ d V ( u ) }{du}   
                         \right|_{u=r_{ik}}
\frac{     \nvec{r}_k   - \nvec{r}_i    }{r_{ik}}
\label{two_body_grads} 
\end{equation}
The symbol 
\begin{equation}
\left. \frac{\tmop{d} V \left( u \right)}{\tmop{d} u} \right| _{u=R_{mn}}
\label{Define_derivative}
\end{equation}
denotes the $(m,n)$ element of the matrix obtained by applying function 
$$
\frac{d V(u)}
   {d u} 
$$ 
on the matrix $R$. We will generalize this notation to any type of objects,
e.g.
\begin{equation}
\left. \frac{d V \left( u \right)}{d u} \right| _{u=Q_{ijk}}
\label{Define_derivative_3}
\end{equation}
denotes the $(i,j,k)$ element of the corresponding 3-index array.  
Without the indices, the discussed symbol  represents  the whole matrix or array.


\section{MATLAB implementation  \label{matlab_2body}}
We shall now show in detail how solutions of eq. \ref{two_body_grads}  can be implemented in very short codes in MATLAB.

For one-dimensional world, to generate an object holding all the distances $d_{ij}=x_j - x_i $,
we can proceed as follows. We can aim at a matrix D with elements $d_{ij}$. In traditional programming
such a construction would require two loops. In MATLAB this can be done in one statement. Provided that
the positions are in a vector $ x=(x_1, x_2...... x_N ) $   
\parskip 0cm 
\parindent 0cm
{  \small  \begin{verbatim}
D=ones(N,1)*x-x'*ones(1,N);
\end{verbatim}        }
In mathematical notation this means
\parskip 0cm 
\parindent 0.95cm
\begin{equation}
\left[
\begin{array}{c}
 1 \\
 1 \\
 1 \\
 .     \\
 .     \\
 1
\end{array}
\right]
          \begin{array}{cccccc}
          \left[ 
           \ \  x_{1} \right. &
                x_{2}         &
                x_{3}         &
              ...             &
          \left. x_{N} \ \  \right] \\
          \ \ & \ \ &  \ \ & \ \ \\
          \ \ & \ \ &  \ \ & \ \ \\
          \ \ & \ \ &  \ \ & \ \ \\
          \ \ & \ \ &  \ \ & \ \ \\
          \ \ & \ \ &  \ \ & \ \ 
          \end{array}
     \ \ \ -  \ \ \  
                    \left[
                    \begin{array}{c}
                    x_{1} \\
                    x_{2} \\
                    x_{3} \\
                    .     \\
                    .     \\
                    x_{N}
                    \end{array}
                    \right]
          \begin{array}{cccccc}
          \left[ 
           \ \   1 \right. &
                1         &
                1         &
                ...             &
          \left. 1 \ \  \right] \\
          \ \ & \ \ &  \ \ & \ \ \\
          \ \ & \ \ &  \ \ & \ \ \\
          \ \ & \ \ &  \ \ & \ \ \\
          \ \ & \ \ &  \ \ & \ \ \\
          \ \ & \ \ &  \ \ & \ \ 
          \end{array}
          \label{Vectors_Distances}
\end{equation}
%
%
%
        and the following matrix operation 
\begin{equation}
 {D}^{x}   
 =
\left[ \begin{array}{c c c c c}
x_{1} &  x_{2} &  x_{3} & \cdots &  x_{\!N} \\
x_{1} &  x_{2} &  x_{3} & \cdots &  x_{\!N} \\
x_{1} &  x_{2} &  x_{3} & \cdots &  x_{\!N} \\
\vdots         & \vdots	         &    \vdots	   & \ddots & \vdots		\\
x_{1} &  x_{2} &  x_{3} & \cdots &  x_{\!N} 
\end{array}\right]
-
\left[ \begin{array}{c c c c c}
x_{1} & x_{1} & x_{1} & \cdots & x_{1} \\
x_{2} & x_{2} & x_{2} & \cdots & x_{2} \\
x_{3} & x_{3} & x_{3} & \cdots & x_{3} \\
\vdots           & \vdots           & \vdots           & \ddots & \vdots           \\
x_{\!N} & x_{\!N} & x_{\!N} & \cdots & x_{\!N} 
\end{array}\right]
\label{doublets_dist} 
\end{equation}
results in the matrix of all scalar distances whose elements are the distances 
$$
D^x_{ij}=x_j-x_i
$$
The diagonal of the matrix D is clearly zero, and that could cause some problems
in inversions and similar operations. Besides, the forces "on itself" are not considered,
so that there should be a mechanism of keeping the diagonal out of the calculations.
The simple mechanism in MATLAB is to use "not" operation on the unity matrix, given
by the built-in function {\tt eye(N) }
\parskip 0cm 
\parindent 0cm
{  \small  \begin{verbatim}
U= eye(N)
\end{verbatim}        }
The matrix Un is $N\times N$ matrix with one 
\parskip 0cm 
\parindent 0.95cm
everywhere except on diagonal, obtained by
\parskip 0cm 
\parindent 0cm
{  \small  \begin{verbatim}
Un = ~eye(N)
\end{verbatim}        }
In MATLAB  there are two types of multiplication: 
the mathematical product of matrices
{  \small  \begin{verbatim}
A * B
\end{verbatim}        }
and the element by element, $ (A \cdot * B )_{ij} \ = \ A_i B_j $, appearing in code as
{  \small  \begin{verbatim}
A.*B
\end{verbatim}        }
It should be rather clear that this can be used as a decomposition
by projection operators into a part which has zeros on the diagonal
and the diagonal part itself:
{  \small  \begin{verbatim}
C= Un .* A  + U .* A;  C= (~U) .* A + U.*A; C= ~eye(N).*A + eye(N).*A
\end{verbatim}        }
In the above code, the three assignments perform the same operation, assigning the
matrix A to C, but the first term in the assignment is always the non-diagonal part.
\parskip 0cm 
\parindent 0.95cm
This decomposition can be used in various combinations, as we shall see below. 
        
The code for our molecular dynamics engine has only one disadvantage: for speed it is
necessary to write the function for the force explicitly, but in principle it could be
specified by a call to function script. Here we shall use the explicit form.

The whole code for the time development of the system of particles, i.e. the
whole molecular dynamics engine in this approach is
\parskip 0cm 
\parindent 0cm
{\small
\begin{verbatim}
 1. Umat=zeros(2,N,N); Umat(1,:,:)=~eye(N,N); Umat(2,:,:)=~eye(N,N);
 2. x=zeros(1,N); y=x; F=zeros(2,N); RR=zeros(2,N,N);
 3. N=input('No of particles > ');  ntime=input('No of time steps > ');
 4. R=zeros(2,N,ntime);     t=0:dtime:ntime*dtime;
 5. start_solutions(R,N,ntime);
 6.
 7. for m = 2:ntime
 8.    x(:)=R(1,:,m);    y(:)=R(2,:,m);
 9.    dx= ones(N,1)*x - x' * ones(1,N) ;
10.    dy= ones(N,1)*y - y' * ones(1,N) ;
11.    dr=sqrt(dx.^2 + dy.^2) + eye(N);  % eye(N) avoids div. by zero
12.    RR(1,:,:)=dr;    RR(2,:,:)=dr;
13.    dR(1,:,:)=dx;    dR(2,:,:)=dy;
14.    forces =  (-a*A*exp(-a*RR) +b*B*exp(-b*RR) ).* ( - dR ./ RR);
15.    F(:,:)=sum(forces.*Umat,3);
16.    R(:,:,m+1)= 2*R(:,:,m)- R(:,:,m-1) + F/M *( t(m)-t(m-1) )^2;  
17. end
\end{verbatim}
}
Note that the above is not a pseudocode, but actual code. 
\parskip 0cm 
\parindent 0.95cm
The fifth line
is a shorthand notation for starting the solution. The Verlet method must 
be started by assigning the starting time values of positions and the volues
at the first time step. Otherwise, the first 5 lines effectively declare the
variables. The 3-index array of positions storing the whole simulation
is R - for all times, the value of x and y (and z in 3-dimensional case)
are the present positions, evaluated in previous step, are equivalently written
as
\parskip 0.1cm 
\parindent 0cm
{  \small  \begin{verbatim}
R(1:2,1:N,1:ntime)       x(1:N)      y(1:N)
\end{verbatim}        }
i.e. specifying the whole definition range gives the object itself. 
\parskip 0cm 
\parindent 0.95cm
We have chosen to present the case of particles moving in 2-dimensional space. 
The 3-dimensional version
of this code is obtained by only adding relevant analogues 
for z, dz and changing
the value 2 to 3 in all the \verb!zeros(2,....)! statements. It is naturally possible
to write the code so that it includes both two and three-dimensional versions, 
but we present here the more transparent explicit form used in the listing.
The actual loop over the time mesh is shown in lines 7-17. Line 8 assigns
the present time values. The matrices \verb!dx! and  \verb!dy!  are discussed above,
they hold the components of the distances between the particles.
We use multidimensional arrays to perform all the operations, therefore there 
is a new complexity element. Since the operations on the forces components
are evaluated in one statement, line 14, the \verb!dr! is duplicated 
into the appropriately dimensioned \verb!RR(1:2,1:N,1:N)!

For  the understanding of the more compact code which we shall use further on we reformulate 
the code in the above example. First we repeat the relevant part to be changed
\parskip 0cm 
\parindent 0cm
{\small
\begin{verbatim}
 8.    x(:)=R(1,:,m);    y(:)=R(2,:,m);
 9.    dx= ones(N,1)*x - x' * ones(1,N) ;
10.    dy= ones(N,1)*y - y' * ones(1,N) ;
11.    dr=sqrt(dx.^2 + dy.^2) + eye(N);  % eye(N) avoids div. by zero
12.    RR(1,:,:)=dr;    RR(2,:,:)=dr;
13.    dR(1,:,:)=dx;    dR(2,:,:)=dy;
14.    forces =  (-a*A*exp(-a*RR) +b*B*exp(-b*RR) ).* ( - dR ./ RR);
\end{verbatim}
}
and below we see the more compact code. This way of coding is now applicable
for both two-dimensional and three-dimensional physical space. First
the instantaneous array of vectors is assigned to r (replacing x and y).
Then the vector multiplications are replaced by constructions explained in the
appendix. 
{\small
\begin{verbatim}
 8.    r(:,:)=R(:,:,m);   
 9.    dR=r(:,:,ones(1,N)) - permute( r(:,:,ones(1,N)) , [1,3,2] );
10.
11.    RR(1,:,:)=sqrt( sum(dR.^2,1 )) + ~Umat ;  % ~Umat avoids div. by zero
12.    RR=RR([1 1],:,:);
13.    
14.    forces =  (-a*A*exp(-a*RR) +b*B*exp(-b*RR) ).* ( - dR ./ RR);
15.    F(:,:)=sum(forces.*Umat,3);
16.    R(:,:,m+1)= 2*R(:,:,m)- R(:,:,m-1) + F/M *( t(m)-t(m-1) )^2;  
17. end
\end{verbatim}
}
The operation permute is a generalization of the transpose operation,
where the permutation of indices is specified by the second variable of the
operation, i.e. if \verb! A=ones(N,N)!, both of the following calls 
{  \small  \begin{verbatim} 
 permute( A, [2,1] )-transpose(A)
 permute( A, [1,2] )-A
\end{verbatim}        }
would return zero 
\parskip 0cm 
\parindent 0.95cm
matrices.

In the last code there is still too much calculation going on.
It can be simplified by having the forces predefined
\parskip 0cm 
\parindent 0cm
{\small
\begin{verbatim}
 8.    r(:,:)=R(:,:,m);   
 9.    dR=r(:,:,ones(1,N)) - permute( r(:,:,ones(1,N)) , [1,3,2] );
10.    rr(:,:)=sqrt( sum(dR.^2,1 )) + eye(N);
11.    RR(1,:,:)=rr;    RR=RR([1 1],:,:);
12.    forces(1,:,:)=  (-a*A*exp(-a*rr) +b*B*exp(-b*rr) );
13.    forces=forces([ 1 1],:,:);
14.    forces =  forces.* ( - dR ./ RR);
15.    F(:,:)=sum(forces.*Umat,3);
16.    R(:,:,m+1)= 2*R(:,:,m)- R(:,:,m-1) + F/M *( t(m)-t(m-1) )^2;  
17. end
\end{verbatim}
}
The calculation might become faster for very large number of particles
but the code in the previous listing is more understandable. 
\parskip 0cm 
\parindent 0.95cm
For small numbers of particles the first formulation might even
perform faster in some cases, simply because the code is shorter.

\subsection{Two body forces reformulated  }      

In this part we reformulate the two-body problem in order to
generalize to the n-body case.
The total energy is a sum over the pairs
\begin{equation}
 V_{tot} = \sum_{i > j} V_{ij} 
\label{sum_V_ij}
\end{equation}
which can be rewritten as, ( provided that $V_{ij} = V_{ji}$ \
which is always satisfied )
\begin{equation}
 V_{\tmop{tot}} = \frac{1}{2} \sum_i \sum_{j \neq i} V_{ij}  
\label{sum_V_i_and_J}
\end{equation}
and the condition $j \neq i$ is automatically reflected by the
properties of the interactions. The force on $m$-th particle is obtained by
\begin{equation}
 \nabla_m V_{\tmop{tot}} = \frac{1}{2} \sum_i \sum_j \nabla_m V_{ij}
\label{gradien_on_sum}
\end{equation}
In evaluating each of the contributions in the sum
\begin{equation}
 \nabla_m V_{ij} =\left.  \frac{d V(u)}{d u} \right|_{u=r_{ij}} \left(
   \nabla_m r_{ij} \right)  
\label{Derivative_V_def_again}
\end{equation}
we must realize that (note that this is the simplest notation, already
including symmetry)
\begin{equation}
 \nabla_m r_{ij} = \frac{\nvec{r}_m - \nvec{r}_j}{r_{ij}} \delta_{im}
   + \frac{\nvec{r}_m - \nvec{r}_i }{r_{ij}} \delta_{jm}  
\label{Kronnecker_delas}
\end{equation}
Due to the properties of the $\delta$-symbols this can be written as
\begin{equation}
 \nabla_m V_{ij} = \left. \frac{\tmop{dV}}{\tmop{dr}} \right|_{ij} \left(
   \nvec{e}_{ij} \delta_{im} + \nvec{e}_{ij} \delta_{jm} \right)
\label{unit_vectors_e_gradient}
\end{equation}
where $\nvec{e}_{ij}$ represents the unit vector in the
direction $\nvec{r}_{ij}$ and the terms after summation will become
$\nvec{e}_{mj}$ and \ $\nvec{e}_{im}$. Since the index m appears at different
positions, after the summations are performed, the two terms will  correspond to
two different positions of the matrix. For simplification of the notation we
will just consider the notation \ $\nvec{e} = (X, Y, Z)$ for the x,y,z
components of the unit vectors  and consider now just the x-components
$X_{ij}$;  the whole matrix will be then denoted by $X$. Thus
\begin{equation}
\frac{1}{2} \sum_i \sum_j  \frac     {\partial}{\partial x_m}        V_{ij} 
=
   \frac{1}{2} 
   \left[     \sum_j 
   \left. \frac{d V \left( u \right)}{d u} \right| _{u=r_{mj}}      
   \text{$X_{mj}$} + 
   \sum_i X_{\tmop{mi}}
   \left. \frac{d V \left( u \right)}{d u} \right| _{u=r_{im}}
   \right]  
\label{x-part-of-gradient}
\end{equation}
We observe that the right hand side of the last eq. \ref{x-part-of-gradient} is in the form of matrix product, in the first
term we use the fact that
$$
   \left. \frac{d V \left( u \right)}{d u} \right| _{u=r_{mj}}     = 
  \left. \frac{d V \left( u \right)}{d u} \right| _{u=r_{jm}}   \ \ 
  \longrightarrow  \ \  \left[ \frac{d V}{d r }  \right] _{jm}
$$
since $r_{mj} = r_{jm}$ and the two terms can be brought to the
same form as the second,
\begin{equation}
 \frac{1}{2} \sum_i \sum_j \frac{\partial}{\partial x_m}_{} V_{ij} =
   \left( X \left[ \frac{\tmop{dV}}{\tmop{dr}} \right]  \right)_{\tmop{mm}} 
\label{two-body-term-reduced}
\end{equation}
i.e. the matrix product of matrix X and the derivative matrix, 
not element-wise product encountered before.
The matrix notation conveniently denotes the result, but it is not needed for
the practical evaluation (it takes $N$ times longer time). The element-wise product of the two matrices (the
second transposed) and summation over the second index will perform the same
operation. In MATLAB notation,
\parskip 0cm
\parindent 0cm
{  \small  \begin{verbatim} 
A= B*C;                 % A matrix N x N   VEC(k) = A(k,k) 
VEC=sum( B .* C.', 2 );    
\end{verbatim}        } 
The reduction to a vector shown in eq. \ref{two-body-term-reduced} 
can be seen as a general rule
which will be useful below, in the three-body case. This formulation
is already implemented in the listings in section \ref{matlab_2body}.
%
%
\parskip 0cm
\parindent 0.95cm

\section{Three body interactions}

   Consider now a system of N mass points whose interactions are described by both
pairwise distance dependent forces and additional three-body forces with dependence
on the angles between the "bonds" to the neighboring points, a situation usually found 
in molecular dynamics.
The two-body part of the forces can be treated as above. To evaluate quantities related 
to triplets of objects, as e.g.  angles between
vectors connecting the particles (known as bond angles), a set of all pairs would 
be first computed, then a loop would run over that set, effectively including a triple 
or even four-fold loop, depending on algorithms chosen. All further evaluations of 
various functions of distances and bond-angles would involve writing efficient loops
for manipulations, evaluations and summations of energies and forces, and solving
the differential equations for the coupled system. Here we discuss the extension
of the approach discussed above to cover also the three-body case. We show the methods
on the example of Stillinger and Weber three-body potentials.
We show that also for this case the operations can be programmed in MATLAB
in simple scalar-like notation and 
from the point of view of efficiency, the mathematical operations 
on matrices,or even multidimensional arrays, can be performed with speeds comparable to 
compiled and optimized code.

\subsection{Stillinger-Weber potential \label{SW_defs}}

A typical three-body potential is the
Stillinger and Weber  potential defined (additional details not important for 
the present discussion
can be found in the original reference) as
\begin{equation}
V_{SW}(\nvec{r}_1,\nvec{r}_2,\nvec{r}_3,....,\nvec{r}_N)
=
\sum_{i<j} V_2  \left( \left| \nvec{r}_1 - \nvec{r}_j  \right| \right)         
+
\sum_{i<j<k} V_3(\nvec{r}_i,\nvec{r}_j,\nvec{r}_k) 
\label{stillinger_TOT_PHI}
\end{equation}
and the force on each particle is still obtained by gradient operation
\begin{equation}
m_k \ddot{\nvec{r}_k}  =-\nabla_k V_{tot}( \nvec{r}_1, ..... , \nvec{r}_N)
\label{3_body_grad} 
\end{equation}
exactly as in eq. \ref{two_body_grads}, but now the simple 
evaluation resulting in sum of forces from the individual neighbours
can not be used as in the second part of eq. \ref{two_body_grads}.

The $V_2(r_{ij})$ part contributes to the forces as described in the previous section.
The new three body part is by Stillinger and Weber chosen as
\begin{equation}
V_3(\nvec{r}_i, \nvec{r}_j, \nvec{r}_k,)=
       h(r_{ij},r_{ik},\theta_{jik})
     + h(r_{ji},r_{jk},\theta_{ijk})
     + h(r_{ki},r_{kj},\theta_{ikj})
\label{stillinger_f_3}
\end{equation}
where $\theta_{ijk}$ is the angle between $\nvec{r}_{ji}$ and $\nvec{r}_{ki}$ 
\begin{equation}
\cos \theta_{ikj}  =  
        \frac{\nvec{r}_k - \nvec{r}_i }{\left| \nvec{r}_k - \nvec{r}_i  \right| } 
        \cdot  
        \frac{\nvec{r}_j - \nvec{r}_i }{\left| \nvec{r}_j - \nvec{r}_i  \right| }  
        \ \ \ \ \ \ \ \ \ \ \ \ \          
        \cos \theta_{ikj}  =  \nvec{e}_{ij}   \cdot   \nvec{e}_{ik} 
\label{stillinger_THETA}
\end{equation}
at the position  of the $i$-th particle. We also introduce here the notation for the
unit vectors  $\nvec{e}_{ij}$ in direction of the distance vectors  $\nvec{r}_{ij}=\nvec{r}_j - \nvec{r}_i $. 
The functions $h(r_1,r_2,\theta)$ are chosen by Stillinger and Weber as
\begin{equation}
h(r_{ij},r_{ik},\theta_{jik}) = 
              g(r_{ij})       
              g(r_{ik})     
            \left(  \cos \theta_{jik}  + \frac{1}{3}     \right)^2
\label{stillinger_h_func}
\end{equation}
where the functions $g(r_{ij})$ mainly play a role of cut-off functions. 
We see that the most important feature of three-body part are the angles 
between the lines connecting the pairs of particles. This is the element
which is going to be investigated in detail in the following
section.

For later use we denote
\begin{equation}
V_{ijk}   =  
h(r_{ij},r_{ik},\theta_{jik}) 
\label{stillinger_V_ijk}
\end{equation}
a notation useful in discussion of the evaluation of the forces below. Note that in this
part we have followed to high degree the notation used by the authors
in the original paper. Below we shall use our notation with stress on the 
pair-separability of this interaction, as discussed below. The main feature is the
 rather simple rewriting by the secongd part of eq. \ref{stillinger_THETA}.

\subsection{Array operations for three-body quantities \label{SW_arrays}}

In eq. \ref{doublets_dist}  we have introduced the fast way to compute and represent 
the distances between all particles - and in the following used both the scalar distances
and their vector versions. These quantities have two indices and are thus conveniently 
treated as matrices, or $N \times N$ objects. When we wished to keep all the space components 
in the same object, we ended with $3  \times N \times N$ array objects.
For three body interactions the potential terms $ V_{ijk} $ of  eq. \ref{stillinger_V_ijk}
are at the outset $N  \times N \times N$ array objects.

\begin{figure}[htp]
\centering%
\includegraphics[ width=9cm]{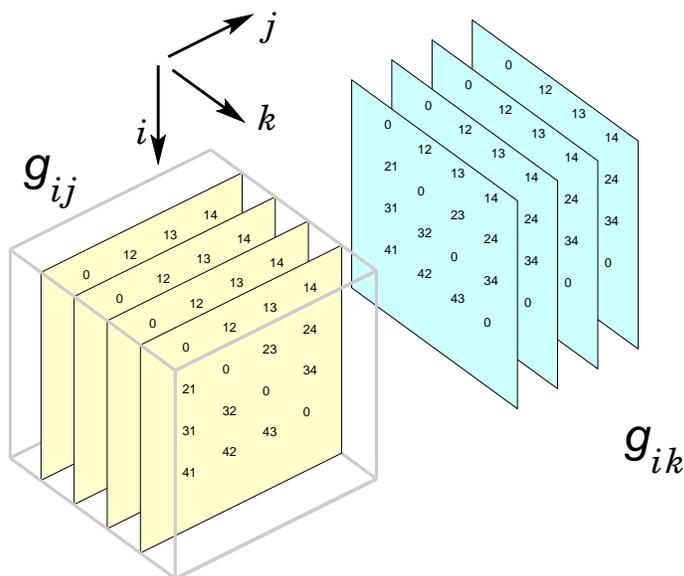}
\caption{The evaluation of triple-index arrays of type $V_{ijk}=g_{ij} g_{ik}$ for three-body forces. 
The matrix with elements $g_{ij}$  is repeated along the 
third index $k$ - forming the "pages" of three-index array. 
The same matrix is duplicated in the direction of the second index, filling the three-index
array in another way. The total three-index array with elements $V_{ijk}$ is obtained
in MATLAB by element-wise multiplication of the two "cubes" shown in the figure. The direction
of the indeces is indicated. This figure illustrates the MATLAB treatment of 
eq. \ref{V_ijk__g_ik}.}
\label{r_ij*r_ik}
\end{figure}

To start we imagine a simplified term: instead of eq. \ref{stillinger_h_func} 
( which we have already rewritten according to a more transparent notation of
eq. \ref{stillinger_THETA} )
\begin{equation}
V_{ijk}  = 
              g(r_{ij})       
              g(r_{ik})     
            \left( \frac{1}{3} +
              \nvec{e}_{ij} \cdot \nvec{e}_{ik}                \right)^2 
\label{V_ijk__cosTheta}
\end{equation}
we first consider only the first two terms in the product
\begin{equation}
V_{ijk}\ \ \ \rightarrow   \ \ \
              g(r_{ij})       
              g(r_{ik})     
\label{V_ijk__g_ik}
\end{equation}
The objects with indices $i j k$ which would represent   
the $g(r_{ij})$   and   $ g(r_{ik})  $ can be built from
the matrices D as shown schematically in figure \ref{r_ij*r_ik}.
\parskip 0cm 
\parindent 0cm
{  \small  \begin{verbatim}     
D3=zeros(N,N,N);
D3(:,:,:)  =  D2 (:,:,ones(N,1))  
\end{verbatim}        
} 
This corresponds to the first cube in figure \ref{r_ij*r_ik},
and it is going to be multiplied by a similar object obtained by permuting the indices
as shown schematically in this listing:
{  \small  \begin{verbatim}     
D3=zeros(N,N,N);
D3(:,:,:)  =  D2 (:,:,ones(N,1)) ;
G1=gfun(D3);
G2=gfun(  permute ( D3, [ 1 3 2 ]  );
V3=  G1  .*  G2;
\end{verbatim}    } 
where the two matrices are multiplied element-wise.
\parskip 0cm 
\parindent 0.95cm

\subsection{Evaluation of the forces from SW potential\label{SW-evaluation}}      

The total energy is sum over the triplets
\[ V_{\tmop{tot}} = \sum_{i > j > k} V (r_i, r_j, r_k) \]
but now the energies inside of the triplet must be specified. Since
\[ V (r_i, r_j, r_k) = V (r_j, r_i, r_k) = \ldots \ldots = V (r_k, r_j, r_i)
\]
\[ V_{\tmop{tot}} = \frac{1}{6} \sum_i \sum_{j \neq i} \sum_{k \neq i, j} V
   (r_i, r_j, r_k) \]
With most reasonable interaction forms, $V (r_i, r_j, r_k) = V_{\tmop{ijk}} +
V_{\tmop{kij}} + V_{\tmop{jki}}$ where the first index specifies the ``vertex
of the ordered triplet'', in the specific case discussed, the atom where the
bond angle is measured. For such cases the summation can run over all
individual cases and results in
\[ V_{\tmop{tot}} = \frac{1}{2}  \sum_i \sum_{j \neq i} \sum_{k \neq i, j}
   V_{\tmop{ijk}} \]
where now only factor 1/2 comes from the fact $V_{\tmop{ijk}} = V_{\tmop{ikj}}$, i.e.
the vertex $i$ is the same but the remaining pair $(\tmop{jk}$) can be counted
twice. The evaluation of the forces is in principle identical, but can become
quite different for cases where different degrees of simplification are
possible.

For The SW case, the interactions can be described as pair-separable. For a
given $V_{\tmop{ijk}}$ one has the form
\begin{equation}
V_{\tmop{ijk}} = g (r_{ij}) g (r_{\tmop{ik}}) 
    f (      \nvec{e}_{ij} \cdot \nvec{e}_{ik}       ) 
   = g (r_{ij}) g (r_{\tmop{ik}}) \left( \frac{1}{3} +
   \nvec{e}_{ij} \cdot \nvec{e}_{ik}         \right)^2  
\label{SW_V_ijk_Defs}
\end{equation}
Clearly, this form of interaction can be built as a three-index array,
produced by element-wise multiplications of arrays constructed by duplication
of matrices. Consider first a simpler form, where we keep just the funcions of
distances,
\[ V^d_{\tmop{ijk}} = g (r_{ij}) g (r_{\tmop{ik}}) \]
The total energy three-index array
$$ 
\left[V^d\right] =  
 {\cal R}\left[ \left( g (r)\right),(\ast ,\ast,N) \right] 
       \cdot \ast 
       {\cal R}\left[ \left(g (r)\right), (\ast ,N,\ast) \right]   
$$
The notation is schematic, and shows \ that the total array $V^d_{}
\tmop{can}$ be constructed as a product of two three index arrays each of which
originates from replication of the original matrices. 
We introduce a symbol  ${\cal R}\left[ \left( A\right),(\ast ,\ast,N) \right]$ 
to denote the replication of matrix $A$ along the dimension which contains the
number, while the dimensions marked by asterix appear once. Construction 
of the two arrays is shown in figure \ref{r_ij*r_ik}.
We have also used the operator $A \cdot
\ast B$ for element-wise multiplication of two arrays of the same dimensions
in analogy with the MATLAB notation.          

The importance of this is that also the gradient operations on such object
will be obtained by operations on the original matrices, i.e. as $\nabla_m g (r_{ij}) $.
These are in form identical to the $\nabla_m
V_{ij} $ and  so will be the $\nabla_m g_{\tmop{ik}}$. These gradients
thus need to be calculated only for the $N \times N$ matrix and
replicated. Schematically we can write, following the above notation,
\begin{eqnarray}
 \nabla_m \left[V^d\right]_{ijk} &  =  &
    \  \  {\cal R}\left[ \left(\nabla_m  g (r)\right),(\ast ,\ast,N) \right]
 \cdot \ast 
   {\cal R}\left[ \left(g (r)\right), (\ast ,N,\ast) \right]  \nonumber \\
   &  \   & + \  
   {\cal R}\left[ \left( g (r)\right),(\ast ,\ast,N) \right] 
    \cdot \ast    
       {\cal R}\left[ \left(\nabla_m g (r)\right), (\ast ,N,\ast) \right] 
\end{eqnarray}
Rewriting of these relations in the MATLAB notation will immediately produce
the desired code. There will be a summation over two indices to produce the
vector of forces from the three-index arrays. The evaluation of the gradients
of the full SW-interaction follows in a straightforward manner the rules of
chain derivative. The details and illustration of the evaluation of the
gradient of the cosine is given in the appendix. Implementing the result shown
in the appendix, we can also there apply the discussed fact that the term
consists of pair-separable terms. Thus all the operations can be also there
performed on matrices of distances and duplicated to the necessary three-index
objects in precisely the same way as shown above for the distance functions.
The only difference is that these are now Euclidian three dimensional vectors
in addition to the particle-index dimensions. Thus the three-body forces can
be constructed in a very similar way as the two-body forces. The computation
is longer only due to the duplication of the matrices and element-wise
products of the resulting matrices. The evaluation of the
functional dependence does not change, since also here the computations are
performed on matrices.


%
\subsection{Symmetry aspects of 3-body force computations \label{COF}}
 The computations above can also be based on symmetry analysis.
For a 3-body potential,  each unique 
interacting triplet of particles has six different permutations contributing thus a total of six forces per index (particle).  
Since the Stillinger-Weber potential -formula  $ V\!\!_{ij k} $, 
defined in eq. \ref{stillinger_V_ijk} -  
exhibits symmetry over the permutation of the second
and third indices, i.e.
 \( V\!\!_{ij k}=V\!\!_{i kj}\), so that forces on the $k$ partilce satisfy
\begin{equation}\label{force1} 
 F\!\!_{ij (k)}=F\!\!_{i (k)j}\;, \hspace {0.5 cm}
 F\!\!_{(k)ij}=F\!\!_{(k)ji} \;,\hspace {0.5 cm}
 F\!\!_{ji (k)}=F\!\!_{j (k)i}
\end{equation}
where the parenthesis denotes the particle coordinate of the gradient
\begin{equation}
F\!\!_{ij (k)}=- \nabla_{k}\!\left(V_{i j k}\right)
\label{force1def} 
\end{equation}
Relations in eq. \ref{force1} demonstrate the symmetry of the problem and show that
only three unique elements of the force array need to be computed for 
each particle in each triplet, as discussed also
in other applications \cite{p105-sumanth},  \cite{ForceDecompositionAlgorithm}. 
Potential formula exhibits also the following antisymmetry due to the triplet's geometrical configuration;
\begin{equation}
 \nabla_{i}\!\left(V_{i j k}\right)=
-\nabla_{j}\!\left(V_{i j k}\right) 
-\nabla_{k}\!\left(V_{i j k}\right)
\label{anti_symmetry_forces}
\end{equation}
which can also be written as 
\(  
F\!\!_{(i)j k}=-\left(F\!\!_{i(j)k} + F\!\!_{ij(k)}\right)
\).
Thus only two unique forces per particle per unique triplet are to be evaluated.
\subsubsection*{Computations of force arrays}
Evaluation of analytical derivatives (gradients) of the 3-body term of the given potential with respect to the 
third index (along the fourth dimension), gives the array, which is denoted \(\mathbbm{F}^{3}\), holding the following six unique 
(two on each particle) force elements per triplet: 
%
\[
\begin{array}{|l|c|r|}
 \hline
 F\!\!_{ij(k)} &F\!\!_{ki(j)} & F\!\!_{ji(k)}\\
 \hline
 F\!\!_{i k(j)}&F\!\!_{kj(i)} & F\!\!_{j k(i)}\\
\hline
\end{array}\]
Those six unique force elements per triplet are what is needed to generate all the other elements by symmetry and antisymmetry exploitation.
Transposing dimensions corresponding to the second and third indices of particles in \(\mathbbm{F}^{3}\) array 
yields a new array \(\mathbbm{F}^{2}\), holding the other six force elements per triplet 
%
\[
\begin{array}{|l|c|r|}
 \hline
 F\!\!_{i(j)k} &F\!\!_{k(i)j} & F\!\!_{j(i)k}\\
 \hline
 F\!\!_{i(k)j}&F\!\!_{k(j)i} & F\!\!_{j (k)i}\\
\hline
\end{array}\]
The third array \(\mathbbm{ F}^{1}\) holding the last six force elements per triplet 
(cf. eq. \ref{anti_symmetry_forces})
%
%
%
\[
\begin{array}{|lllll|}
 \hline
 F\!\!_{(i)j k}&=& - \left(F\!\!_{ij(k)}+F\!\!_{i(j)k}\right)&=&F\!\!_{(i)kj}\\
 \hline
 F\!\!_{(k)ij}&=& -  \left(F\!\!_{ki(j)}+F\!\!_{k(i)j}\right)&=&F\!\!_{(k)ji}\\
\hline
F\!\!_{(j)i k}&=&  - \left(F\!\!_{ji(k)}+F\!\!_{j(i)k}\right)&=&F\!\!_{(j)ki}\\
\hline
\end{array}\]
is computed by the antisymmetry relation:
\[
\mathbbm{ F}^{1} = -\left(\mathbbm{ F}^{2} + \mathbbm{ F}^{3}\right)
\]
Finally the total force on \(i\)-particle is a double sum  over neighbors in triplets, i.e.
\[\displaystyle
F\!\!_{i} =\sum_{{\substack{j \neq i \\ k \neq i, k\neq j}}}
\left(F\!\!_{(i)j k} + F\!\!_{(i)kj}\right) + 
\left(F\!\!_{j(i) k} + F\!\!_{k(i)j}\right) + 
\left(F\!\!_{j k(i)} + F\!\!_{kj(i)}\right)
\] 
This will be done in array form by addition of arrays \(\mathbbm{ F}^{1}\),
\(\mathbbm{ F}^{2 }\) \verb![1 3 2 4]!-permuted    and \(\mathbbm{ F}^{3 }\)
  \verb![1 4 3 2]!-permuted (the permutation of array dimensions is done to place
  the particle $i$ in the same position, see Appendix). The resulting 4-index array
and then summing over the third and fourth index of results in $3\times N$ array of the forces.


Similar considerations and constructions can be done for the three-index arrays and their gradients
in the case of Tersoff-type potentials  in section \ref{TersoffFormula} below, but there the symmetry 
and antisymmetry occur at two different stages of construction of the forces.
\subsection{Experiments with three-body forces}
In this section we have shown how one can use MATLAB functionality for 
evaluation of three body potentials and forces for three-body interactions. 
The operations on three-index arrays of the type discussed are "expensive" in computer
time but rather straightforward in terms of programming. 
The use of symmetries and more efficient build-up as discussed 
in sections \ref{SW-evaluation} and \ref{COF} makes
the computations much faster, but is much more error-prone. Here
it can be useful to use a simple trick, formulate the computations
first in a slow but easier programmed full-object version, even perhaps
with the Kronnecker-deltas shown in eq. \ref{unit_vectors_e_gradient} implemented
and explicitely summed,  
and then gradually introduce the 
more efficient computations in further versions. This offers possibilities for
code development which are difficult to match in simplicity 
in other approaches. However, this approach offers
even more benefits when investigating
more complex interactions, as the Tersoff type discussed in the following
section.
\section{Tersoff-type many body potential}\label{TersoffFormula}

A more complex many-body potentials quite extensively
used are the Tersoff type potentials. They are constructed in
a slightly different way, but the techniques discussed in previous section
can be used with some modifications also here.
The many body character is introduced via a pair interaction,
but the parameters of the interaction depend on the positions 
of the neigbours and 
the same bond angles for triplets formed by the pair and a neighbour atom.
We first follow the formulation usually found in literature 
and later translate this to our notation.
Considering a Morse-like potential,  
the energy can be written as a sum of cite energies as follows:
\begin{equation} 
V_{tot} =   \frac{1}{2} \sum_{i,j \neq i} V_{ij} ,
\label{tersoff_a} 
\end{equation} 
where
\begin{equation} 
V_{ij} =  \begin{array}{ccc} 
f_c(r_{ij})\Bigl( f_{R}(r_{ij}) & + & b_{ij} f_{A}(r_{ij})\Bigr)
\end{array}
\label{tersoff_b} 
\end{equation}
but where $  b_{ij} =  b (\nvec{r}_{ij};\ \nvec{r}_{1}, \nvec{r}_{2}, \nvec{r}_{3},... \nvec{r}_{N})$, i.e. 
their value is dependent on all the other particles, in practice on those which are in the
neighborhood, as assured by the cut-off function.
The function \(f_c(r_{ij})\) is 
an aproperiate cutoff function, while \(f_{R}\) and \(f_{A} \) are refered to as 
repulsive and attractive pair potentials respectively. 

The two potential functions of the pair-wise interactions are:
\begin{equation} 
\begin{array}{ccc} 
f_{R}(r_{ij}) & = & A \hspace{0.05in} e^{-\lambda_1 r_{ij}} \\
f_{A}(r_{ij}) & = & -B \hspace{0.05in} e^{-\lambda_2 r_{ij}} 
\end{array}
\label{tersoff_c} 
\end{equation} 
The many-particle character in Tersoff's formula is introduced  by the following:
\begin{equation} 
b_{ij}=\frac{1}{(1 + \beta^n {\eta_{i j}}^n)^{\frac{1}{2n}}}
\label{tersoff_d} 
\end{equation}
\[
\eta_{i j}=\sum_{k \neq i, j}\zeta_{i j k}
\]
\begin{equation} 
\zeta_{i j k}= \hspace{0.05in}f_{c}(r_{i k}) \hspace{0.05in} g(\theta_{i j k})\hspace{0.05in}
e^{\lambda_{3}^3(r_{ij}-r_{i k})^3}
\label{tersoff_e} 
\end{equation}
\begin{equation} 
g(\theta_{ij k})= 1 + \frac{c^2}{d^2} - \frac{c^2}{d^2 + \bigl(h-\cos(\theta_{i j k})\bigr)^2}
\label{tersoff_fff} 
\end{equation}
where \(\theta_{ij k}\) is the angle between the       
\(\vec{\bm{r}}_{i j} \) and \(\vec{\bm{r}}_{i k}\) 
as defined above, and the set of parameters for carbon is given in \cite{TersoffCarbon}.
In our work we used a Fermi-type cutoff function which has proven continuity and smoothness (infinitely differentiable) at any point,
\[\displaystyle
f_c(r)= \frac{1}{e^{\left(\frac{r-\mu}{\nu}\right)}+1}
\]
where \(\it {R_1 =1.7,\hspace{.2 cm} R_2 =2.0,\hspace{.2 cm} \mu =0.5(R_2 + R_1),\hspace{.2 cm} \nu =\frac{\pi}{20}(R_2 - R_1)}\), and it  
is defined and parametrized in correspondence to that introduced by Tersoff \cite{TersoffNewEmpirical},
which has a different functional form but a very similar shape. Note that the same cut-off limits the 
pair interaction and the influence on the pair by the neighbors, the same $f_c(r)$
appears both in eq. \ref{tersoff_c} and eq. \ref{tersoff_e}.

To apply methods discussed in the previous section, we realize that the
gradient operation can work on the various parts of the functional dependence.
For simplicity, we omit the cut-off function in this discussion. The force on
each particle $m$ is calculated by applying the the gradient
\begin{equation} 
 \frac{1}{2} \nabla_m \sum_i \sum_{j \neq i} V_{ij}
\label{Nabla_TotalPotential_Tersoff}
\end{equation}
where we now re-write the above eqs. \ref{tersoff_a} to \ref{tersoff_fff} 
in a more transparent notation
\begin{equation} 
V_{ij} = \begin{array}{ccc}
     V_R (r_{ij}) & - & V_A (r_{ij}) B \left( 
     \sum_k 
     g (r_{ij}) g(r_{ik}) 
     d \left( \nvec{e}_{ij} \cdot \nvec{e}_{ik} 
     \right)
     \right)
   \end{array} 
\label{NewNotation_Tersoff}
\end{equation}
or   
\begin{equation}
 V_{ij} =   
     V_R (r_{ij})  -  V_A (r_{ij}) B \left( Z_{_{\tmop{ij}}} \right)
\label{Intro_Z}
\end{equation}
where
\begin{equation}
Z_{ij}  = \sum_k Q_{ijk} =
    \sum_k 
     g (r_{ij}) g(r_{ik}) 
     d \left(     \nvec{e}_{ij} \cdot \nvec{e}_{ik}     \right) 
\label{Intro_Q}
\end{equation}
%
%
%
 %
 %
 %
Applying the gradient,
\begin{equation}
 \frac{1}{2} \nabla_m \sum_i \sum_{j \neq i} V_{ij} = \frac{1}{2} \sum_i
   \sum_{j \neq i} \left( \nabla_m V_R (r_{ij}) - \nabla_m \left[
   \begin{array}{l}
     V_A (r_{ij}) B \left( \sum_k Q_{\tmop{ijk}} \right)
   \end{array} \right] \right) 
\label{Sums_Q_Tersoff}
\end{equation}
we recognize that the first term \ $V_R (r_{ij})$ \ is simply a two-body case
discussed before. The really multi-body term is
\begin{equation}
\nabla_m \left[ \begin{array}{l}
     V_A (r_{ij}) B \left( Z_{\tmop{ij}} \right)
   \end{array} \right] = \nabla_m \left[ \begin{array}{l}
     V_A (r_{ij}) B \left( \sum_k g(r_{\tmop{ij}}) g(r_{\tmop{ik}}) d \left(
     \nvec{e}_{ij} \cdot \nvec{e}_{ik}  \right) \right)
   \end{array} \right] 
\label{The_Tersoff_difference}
\end{equation}
The evaluation of the elements of matrices $V_A (r_{ij})$ and $Z_{ij}$ as
well as the function $B \left( Z_{\tmop{ij}} \right)$ can be done in a
%
%
straightforward way: construct the distance matrix as in  eq \ref{doublets_dist}, 
the unit vectors, and then
construct the three-index array $Q_{\tmop{ijk}}$  as in eq. \ref{stillinger_THETA}, in
MATLAB notation 
\parskip 0.2cm
\parindent 0cm

{ \small \verb!Z=sum(Q,3);  !   }

produces the matrix \verb!Z!. 
\parskip 0cm
\parindent 0.95cm
We can treat all of the terms as before, except the
term
$$
     V_A (r_{ij}) \nabla_m B \left( \sum_k g (r_{\tmop{ij}}) g (r_{\tmop{ik}})
     d \left( \nvec{e}_{ij} \cdot \nvec{e}_{ik} \right) \right)
    = 
     V_A (r_{ij}) \nabla_m B \left( Z_{\tmop{ij}} \right)
   = V_A (r_{ij}) 
\left. \frac{\partial B \left( u \right)}{\partial u} \right| _{u=Z_{ij}}  
   \nabla_m Z_{ij}   
$$  
but we recognize that there is only one new typ of object, \
\begin{equation}
 \nabla_m Z_{ij} = \sum_k \nabla_m Q_{ijk}  
\label{nabla_of_Z}
\end{equation}
but the $\nabla_m Q_{\tmop{ijk}}$ has exactly the same form as the
whole three-body contribution   eq. \ref{SW_V_ijk_Defs} in the SW case and thus can be evaluated
and reduced to a matrix in the same way as in the SW case. The difference is that
it must be combined with the other terms above.
In spite of its complicated
formal nature the Tersoff-type potentials  still contain only quantities including triplets
of particles
and the construction of the code is only a slightly more complex modification 
of the method outlined in section  \ref{SW_arrays}.



\section{Conclusion}
   
The discussed method, implementing multidimensional arrays along with their pre-defined operations, 
has proven to be a useful alternative to the standard loops aggregations in the vector-based MATLAB-like languages. 
It enables fast and powerful memory manipulations  for interactive or exploratory
work. In contrast to explicit loop programming it opens 
for speed-up and efficiency comparable to well optimized compiled  
arithmetic expressions in the framework of an interpreted
code. 
Speed improvement is a main result of vectorized coding to which any numerical algorithm is very 
sensitive. Vectorization is known as performing operations on data stored as vectors, MATLAB stores matrices and arrays 
by columns in contiguous locations in the memory, and allows thus the use of vectorization on those objects.
The elements of arrays can be selectively accessed in a vectorized fashion
using so called logical indices. Mutidimensional logical arrays, as logical indices, have proven to be efficient and 
their use has resulted in a speed gain. 
Finally, the practicality, simplicity and generality of such a method to program empirical formulae for molecular 
dynamics purposes, make it effective to perform simulations on relatively small systems, up to 140 particles in a personal machine
of 1 GB RAM, for Tersoff's potential, a bigger system when using Stillinger-weber potential, and up to 1000 particles if only a pair  
interactions are used. 
The described methodology can be applied  both to investigation of aspects of molecular dynamics
and to educational purposes. Finally, programming in this method results in very 
short codes which are easily used and modified in interactive sessions.  
\appendix
\section*{Appendices}
\section{Some aspects of MATLAB language}\label{AppendixA}

\subsubsection*{  Matrix(Array) manipulation in MATLAB}
\begin{description}
\item{\bf {\em Matrix(array) entering } }
The lines starting with \verb! >> ! are user input, the rest is matlab response. If input is ended
by semicolon, no output follows.     
{  \small  \begin{verbatim}
>>  A=  [  1 4 5 2.5 7 ]    
A
    1.0000    4.0000    5.0000    2.5000    7.0000
>>  B= [  1; 4; 2.5] 
B =
    1.0000
    4.0000
    2.5000
>>  C= [  0 1 ;  1  0]
C =
     0     1
     1     0
>> D=[ 1 4 3 5 2
       3 7 6.6 2 2 ]
D =
    1.0000    4.0000    3.0000    5.0000    2.0000
    3.0000    7.0000    6.6000    2.0000    2.0000
\end{verbatim}        }
The last example, matrix \verb! D ! shows multi-line possibility, the previous is 
a more compact matrix input.
\item{\bf {\em Accessing matrix(array) elements: Basic indexing}}
 
Assume \verb!A! is a random 6 by 6 matrix  \verb!A=rand(6,6)!

\begin{tabular}{l l l }
\verb! A([1 2 3],4)     ! &    returns a column vector     &\verb!  [  A(1,4) ; A(2,4) ; A(3,4) ]  !\\
\verb! A(4,[1,1,1])     ! &    returns a row vector        &\verb!  [  A(4,1) A(4,1) A(4,1)  ]   ! \\
\verb! A([2,5],[3,1])   ! &    returns a 2 by 2 matrix     &\verb!  [  A(2,3)  A(2,1) ; A(5,3)  ! A(5,1) ]
\end{tabular}
            
\item{\bf {\em Matrix(Array) size, length and number of dimensions:}}

The size of an array can be determined by the \verb!size()! command as :
{  \small  \begin{verbatim}
[nrows,ncols, npages, ....] = size(A)
\end{verbatim}        }
where 
{  \small  \begin{verbatim}
nrows=size(A,1),     ncols=size(A,2),     npages=size(A,3)  
\end{verbatim}        }
This function is of much importance when aiming elements-wise operations as it is the key for arrays' size matching.

The 
\verb!ndims(A)! 
    function reports how many dimensions the array \verb!A! has. 
Trailing singleton dimensions are ignored but singleton dimensions occurring before non-singleton dimensions are not. This means that
{  \small  \begin{verbatim}
a=5;   c=a(1,1);  B=zeros(5,5); B(5,5,1)=4;
\end{verbatim}        }
are valid expressions, even though they seemingly have non matching number of indices, but it is 
essential that the extra dimensions are "singletons", i.e. the extra dimensions have length one.

An interesting construction: one can construct matrices and arrays expanding the singleton dimensions.
{  \small  \begin{verbatim}
a=5;   B=a([1 1],[1 1 1 1])
B =
     5     5     5     5
     5     5     5     5
\end{verbatim}        }

The \verb!length()! command gives the number of elements in the longest dimension, and is equivalent to
\verb!max(size())!. It is also used with a row or a column vector.

\item{\bf {\em Matrix(Array) replication:}}

 One can use the MATLAB built-in script repmat, to replicate a matrix by the command:  \verb!repmat(A,[m n])!,
 which replicates the matrix  \verb!A(:,:)! 
 m-times in the first dimension (rows), and n-times in the second dimension (columns). This is a block matrix construction, the block corresponding to the original matrix A is repeated in a m $\times$ n "supermatrix". 
The operation repmat is not a built-in but it is a script which means a slow execution.  
An alternative efficient way to achieve the same replication is by :

\verb!A([1:size(A,1)]'*ones(1,m),[1:size(A,2)]'*ones(1,n))!
which is an application of basic indexing techniques.

In this work we can use another aspect of  \verb!repmat()! script, for building multidimensional arrays
from matrices. Given a matrix A, its elements can be accessed by \verb!A(:,:)!, but that is equivalent to 
\verb!A(:,:,1)!, but also \verb!A(:,:,1,1)!, this is referred to as "trailing singleton dimensions".
Using this, one can replicate the trailing dimensions. From an N$\times$N matrix we can construct a three-index N$\times$N$\times$N array by
\verb!B=repmat(A,[1,1,N]); !  \ \ In the text we discuss a more efficient equivalent
\verb!B=A(:,:,ones(1,N) ) !

\item{\bf{\em Transpose and its generalization permute:}}

For a matrix (2D array) non-conjugate transpose is given by
\verb!transpose(A)! and it is equivalent to a standard matrix operation \verb!A.'!. The normal
Hermite type conjugation is denoted by \verb!A'! and has an equivalent function \verb!ctranspose(A)!.
The \verb!A.'!-notation is somewhat surprising, since the dot usually denotes element-wise 
counterpart of the undotted operator (as e.g. \verb!A.^2!, or \verb!A .* B! etc).
 
 The generalizations of transpose-function to multi-dimensional arrays are 
given as \verb! permute(A,ORDER) ! and \verb! ipermute(A,ORDER) ! 
where ORDER means an array of reordered indeces. When ORDER  is [1 2 3 4] for 4-index array,
it means identity.

\verb!transpose(A)!   and   \verb!permute(A,[2 1])!  give the same  results as (\verb!A.'!). 
The two permute-functions are related as shown in the example

\verb!      H=permute(G,[3,1,2]), G=ipermute(H,[3,1,2])  !

If \verb!size(G) ! is e.g. \verb! 3 4 2 ! then \verb!size(H) ! is  \verb! 2 3 4 !

Clearly, the inverse permute is introduced for convenience, the same result could be achieved 
by using the permute with appropriate order of dimensions.

\item{\bf{\em Logical array addressing and predefined functions:}}

For a given matrix or array \verb!G! the function \verb!logical(G)! 
returns an equal size array where there are logical ones (true) in elements which were nonzero
and logical zeros (false) where the original array had zeros. Thus

\verb!Lg=logical(G)!  and \verb!Lg=G>0!   both return the same matrix of logical ones and zeros.

The extreme usefulness of these functions is that when we write

\verb!G(Lg)= 1  ./ G(Lg) !
the calculation will be performed only for the logically true addresses, thus no division by zero
will happen. Or, we can "shrink" a range of values; for example all values below a threshold
can be given the threshold value and all values above a chosen top value can be given the top value.
As an example, we take a matrix which
should be modified by replacing all values smaller than 2 by 2 and all larger than 5 by 5.
This can be done by the following commands

\verb!      B=A;   B(A>5) = A(A>5)*0+5;   B(A<2) = A(A<2)*0+2  !

The original matrix and the resulting one are 
\parskip 0.0cm 
\parindent 0cm
{  \small  \begin{verbatim}
A=[ 1   5   3   6          B=  2   5   3   5
    7   3   1   1              5   3   2   2
    9   3   4   1              5   3   4   2
    3   4   2   1              3   4   2   2
    9   2   3   1 ]            5   2   3   2
\end{verbatim}        } 
\parskip 0cm 
\parindent 0.95cm
There are also available functions analogous to \verb!ones( )! and \verb!zeros( )!.
As an example, \verb!false(5,2,2)! returns the same logical array as \verb!zeros(5,2,2)>0 ! 
and \verb!true(5,2,2)! is the same as e.g. \verb!logical(ones(5,2,2))!

The use of logical addressing can shorten some large array manipulations
and calculations, but also consumes some processing time.

One can test the gains and losses by filling a large array and doing the
element-wise operations on the whole array and the logically addressed part 
of the array.

In this example we show the array of ten million numbers, where only 100 are nonzero.
{  \small  \begin{verbatim}
N =  10000000;
octave >  tic;  C=zeros(N,1);  C(100:200)=1;  D=exp(C);  toc
     Elapsed time is 0.601 seconds.
     
octave > tic;C=zeros(N,1);C(100:200)=1;    D(C>0)=exp(C(C>0));toc
     Elapsed time is 0.394 seconds.
\end{verbatim}        } 
In the second line we have calculated only the nonzero cases using
the discussed logical addressing.
We observe that there is a certain small gain in time, but far under 
the factor $10^5$ expected from the number of calculations. It should also
be mentioned that the return times are not reproduceable, they depend on the
over all state of the computer memory.
\end{description}
\subsubsection*{Speed improvement in Matlab}
\begin{description}
\item{\bf{\em  Array Preallocation :\\}}
Matlabs matrix variables dynamically augment rows and columns and the matrix is automatically resized. 
Internally, the matrix data memory must be reallocated with larger size. If a matrix is resized repeatedly like within a for loop, 
this overhead becomes noticeable. When a loop is of a must demand, as the inevitable time-loop in MD programs, frequent 
reallocations are avoided by preallocating the array with one of the \verb!zeros()!, \verb!ones()!, 
\verb!rand()!, \verb!true()! 
and \verb!false()! 
commands. 

\item{\bf{\em Vectorization of MATLAB's computation and  logical addressing:}} 
Vectorized computations which mean replacing parallel operations, i.e., simultaneous execution of those operations in loops, with vector operations, improves the speed often ten-fold. For that purpose,
most standard functions in Matlab are vectorized, i.e.,  they operate on an array as if the function was applied individually to every element. 
Logical addressing operations are also vectorized, the use of logical addressing may lead to vectorized computations only for elements where the indices are true, thus reducing execution time, in some cases substantially. 
\end{description}

\section{Gradient of the cosine of the bond angle for three body forces }
\label{GradAngCos}

Let a, b and c mark three atoms, the b-atom is at the vertex of the angle $\theta_{abc}$; \\ 
Define the following quantities: \\
$ \nvec r_{bc}                      = \nvec{r}_c    -  \nvec{r}_b = r_{bc}  \nvec e_{bc} $; \ 
$ \nvec r_{ba}                      = \nvec{r}_a    -  \nvec{r}_b = r_{ba}  \nvec e_{ba}   $; 
$ \cos  \theta_{abc}   = \nvec e_{ba} \cdot  \nvec e_{bc} $;
                        \begin{center}
                                     \includegraphics[ width=5cm]{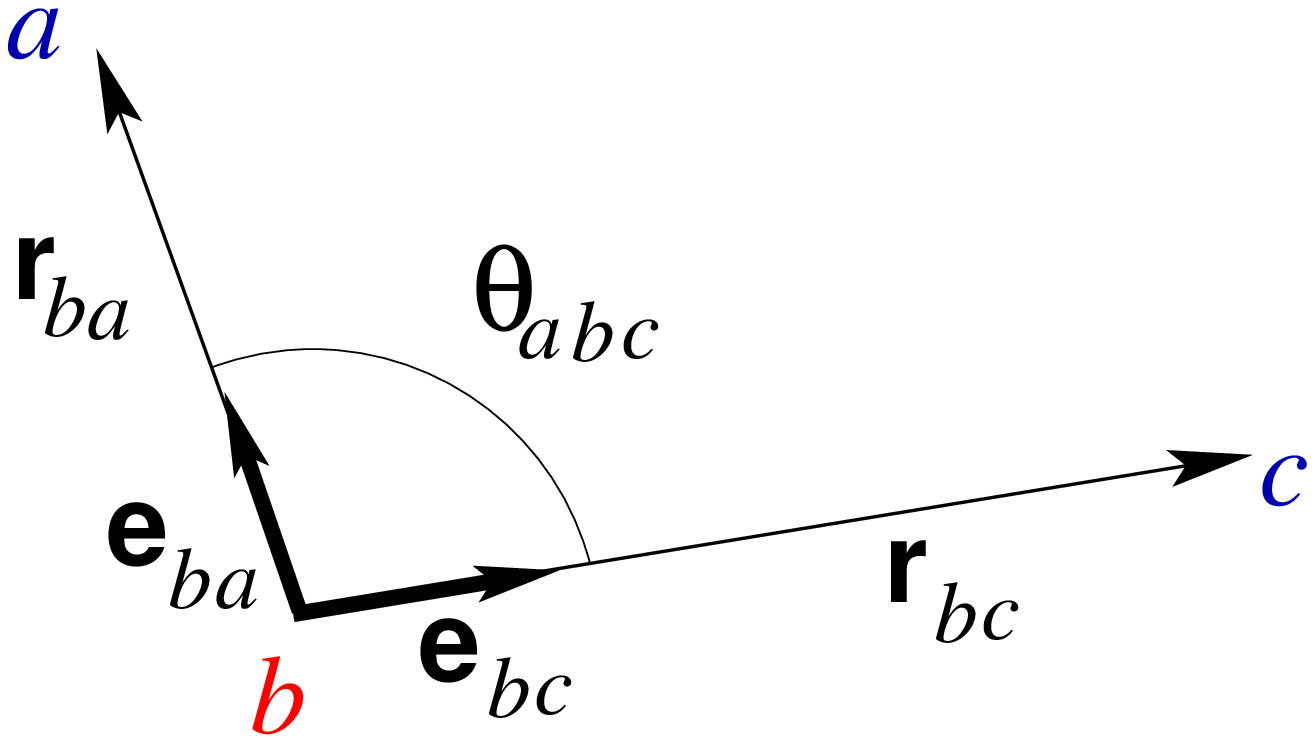}
                         \end{center}

Gradient with respect to the end $c$
\begin{equation}
\nabla_c \cos(\theta_{abc})= 
 \frac{1}{r_{bc}}\bigg[ \nvec e_{ba} - \nvec e_{bc} \cos  \theta_{abc}  \bigg]
\end{equation}
Gradient with respect to the end $a$
\begin{equation}
\nabla_a \cos \theta_{abc}= \frac{1}{r_{ba}}[ \nvec e_{bc} -
\nvec e_{ba} \cos\theta_{abc} ]
\end{equation}
 gradient with respect to point 'b'  :
\begin{equation}
\nabla_b \cos \theta_{abc} = 
- \frac{1}{r_{bc}}\bigg[ \nvec e_{ba} -\nvec e_{bc} \cos \theta_{abc}  \bigg] 
- \frac{1}{r_{ba}}\bigg[ \nvec e_{bc} -\nvec e_{ba} \cos \theta_{abc} \bigg]
\end{equation}
The gradients satisfy the following relation
\begin{equation}
 \nabla_b \cos \theta_{abc}  = - \nabla_a \cos \theta_{abc}  -  \nabla_c \cos \theta_{abc} 
 \label{nabla_nabla_nabla}
\end{equation}
which simply confirms the fact that moving the vertex point $b$ by a vector $\Delta \nvec{r}$ is the same as moving
each of the end points $a$ and $c$ simultaneously by the vector $- \Delta \nvec{r}$.
{\bf Derivation: }

In view of eq. \ref{nabla_nabla_nabla} it is necessary to evaluate just 
one endpoint, the second follows from symmetry
and the vertex is given by relation eq. \ref{nabla_nabla_nabla} 

The calculation is done by using
\begin{equation}
\nabla_r  ( \nvec r \cdot \nvec v_{const} ) =  \nvec v_{const} 
\end{equation}
and 
\begin{equation}
\nabla_r \frac{1}{r} = - \frac{ \nvec r}{r^3 } = - \frac{ 1 }{r^2 }  \nvec e_r 
\end{equation}

Applying these two to the chain derivatives in the gradient as
\begin{equation*}
 \nabla_c  \cos  \theta_{abc}   =   \nabla_c  ( \nvec e_{ba} \cdot  \nvec e_{bc} )
 =  \nabla_c \left( \frac{  \nvec r_{bc}  }{r_{bc}}   \cdot   \frac{  \nvec r_{ba}  }{r_{ba}}   \right)
 \end{equation*}
\begin{equation*}
  \nabla_c \left( \frac{  \nvec r_{bc}  }{r_{bc}}   \cdot   \frac{  \nvec r_{ba}  }{r_{ba}}   \right)
  =  \frac{  1  }{r_{bc}}  \frac{  \nvec r_{ba}  }{r_{ba}}   -   
               \frac{  \nvec r_{bc}  }{r_{bc}^3}  \frac{ \nvec r_{bc}   \cdot   \nvec r_{ba} }{r_{ba}}
= \frac{  1  }{r_{bc}} \left[     \frac{  \nvec r_{ba}  }{r_{ba}}    
                      -  \frac{  \nvec r_{bc}  }{r_{bc}} 
                         \left(   
                         \frac{  \nvec r_{bc}  }{r_{bc}}   \cdot   \frac{  \nvec r_{ba}  }{r_{ba}}      
                         \right) \right]
 \end{equation*}

\nocite{TersoffCarbon,TersoffNewEmpirical} 
\bibliographystyle{plain}

\end{document}